\documentclass[
]{ceurart}

\usepackage{pifont}
\usepackage{xcolor}
\usepackage{graphicx} 
\usepackage{multirow}
\usepackage{caption}
\usepackage{subcaption}

\makeatletter
\@ifundefined{c@rownum}{%
  \let\c@rownum\rownum
}{}
\@ifundefined{therownum}{%
  \def\therownum{\@arabic\rownum}%
}{}
\makeatother

\begin{document}

\copyrightyear{2024}
\copyrightclause{Copyright for this paper by its authors.
  Use permitted under Creative Commons License Attribution 4.0
  International (CC BY 4.0).}

\conference{HCI SI 2023: Human-Computer Interaction Slovenia 2023, January 26, 2024, Maribor, Slovenia}

\title{Retzzles: Do Jigsaw Puzzle Actions on Interactive Display Maps Increase the Retention of Map Information?}


\author[1]{Nikola Kovačević}[%
orcid=0000-0002-7913-3839,
email=kovacevicnikola@proton.me
]
\author[1,3]{Jordan Aiko {Deja}}[%
orcid=0000-0001-9341-6088,
email=	jordan.deja@famnit.upr.si
]
\author[1,4]{Maheshya Weerasinghe}[%
orcid=0000-0003-2691-601X,
email=maheshya.weerasinghe@famnit.upr.si
]
\author[1,2,5]{Klen {Čopič Pucihar}}[%
orcid=0000-0002-7784-1356, 
email=klen.copic@famnit.upr.si,
]
\author[1,5]{Matjaž Kljun}[%
orcid=0000-0002-6988-3046,
email=matjaz.kljun@upr.si,
]

\address[1]{University of Primorska, Faculty of Mathematics, Natural Sciences and Information Technologies, Koper, Slovenia}
\address[2]{Faculty of Information Studies, Novo Mesto, Slovenia}
\address[3]{De La Salle University, Manila, Philippines}
\address[4]{University of Glasgow, Scotland, United Kingdom}
\address[5]{Stellenbosch University, Department of Information Science, Stellenbosch, South Africa}
\begin{abstract}
While maps provide upfront content, this might not always be the most effective way for users to remember information. With the proliferation of interactive displays for tourists and visitors in public spaces, we can create a more playful user experience with maps than just exploring them. Adding interactions with the map could also help users retain more information as they use them. In this paper, we investigated whether completing a jigsaw puzzle of a map supports users in retaining more information about a specific map. The results of a between-subject study with a sample of $n=28$ indicate that additional interaction helped improve mean scores of textual and spatial recall but not visual recall. However, the results are not statistically significant, and the topic is subject to further investigation. Our findings contribute to discussions on using interactive touchscreen displays in similar learning scenarios involving memory retention. 
\end{abstract}


\maketitle

\section{Introduction and Background}

\par Throughout history, maps have been essential resources for exploration, navigation, and communication~\cite{farrell2003transfer}. They are often used in tourism for an overview of points of interest since they support memory and spatial cognition and display information quickly~\cite{jancewicz2017tourist, rossetto2012embodying}. However, traditional paper-based maps frequently fall short in providing further resources as paper real estate is always limited. Additional resources can be delivered either on extra paper-based material or in a digital form~\cite{nakazawa2007phygital}. For example, looking at a physical town map on the board, one can explore additional digital information over the phone. Fully digital maps provide the possibility of interactivity and unlimited further resources, enabling more dynamic user interaction with the content. Another advantage of digital maps is the possibility of showing additional elements in the map's context, which paper-based maps cannot achieve. It has been observed that showing annotations linked to a target object's actual location (that is, right on the map) improves users' memory skills more than showing annotations related to an unrelated location (that is, off the map and connected to their location through other means)~\cite{fujimoto2012relation}. 

\par Digital maps are often provided on interactive displays that are common in tourist information offices, museums, shopping malls, train stations, and even streets~\cite{makela2019supporting}. These displays open up further possibilities for making interaction with maps playful and, in turn, help people remember more information~\cite{ballagas2008gaming, zarzuela2013educational, rooney2013actually, sotlar2020predmetnik, 2019buildingnodalo, iii2021explore}. Playing activity enables the notion of ``learning by doing'' or experiential learning~\cite{weerasinghe2019educational}. There are various ways to make map interaction playful—one possibility is to use jigsaw puzzle pieces~\cite{coltekin2015utilization, matthews2021impact}. Puzzles have been observed to help make learning enjoyable~\cite{keshta2013effectiveness, linehan2014learning}. Specifically, the use of jigsaw elements, one of the most common puzzle games, has been observed to help students acquire a deeper understanding of certain concepts and terminologies~\cite{khalvandi2016effect} in the broader context of learning. 

\par In this paper, we explored whether engaging with maps through playful elements such as jigsaw puzzles can improve engagement and retention of information. We hypothesise that users playing with and completing a jigsaw puzzle of a map will retain more information about it compared to users who will start interacting with the complete map. 
In summary, this research presents the following contributions: a) an improvement of \textbf{Artifact} prototype \textit{Retzzles}~\cite{kovacevic2022retzzles} that is presented on an interactive display, and b) \textbf{Empirical} findings of a user study comparing visual, spatial and textual recall between two conditions --  map interaction vs. jigsaw puzzle interaction.

\section{Retzzles: Design and Implementation}

\begin{figure}[t]
  \centering
  \includegraphics[width=1.0\linewidth]{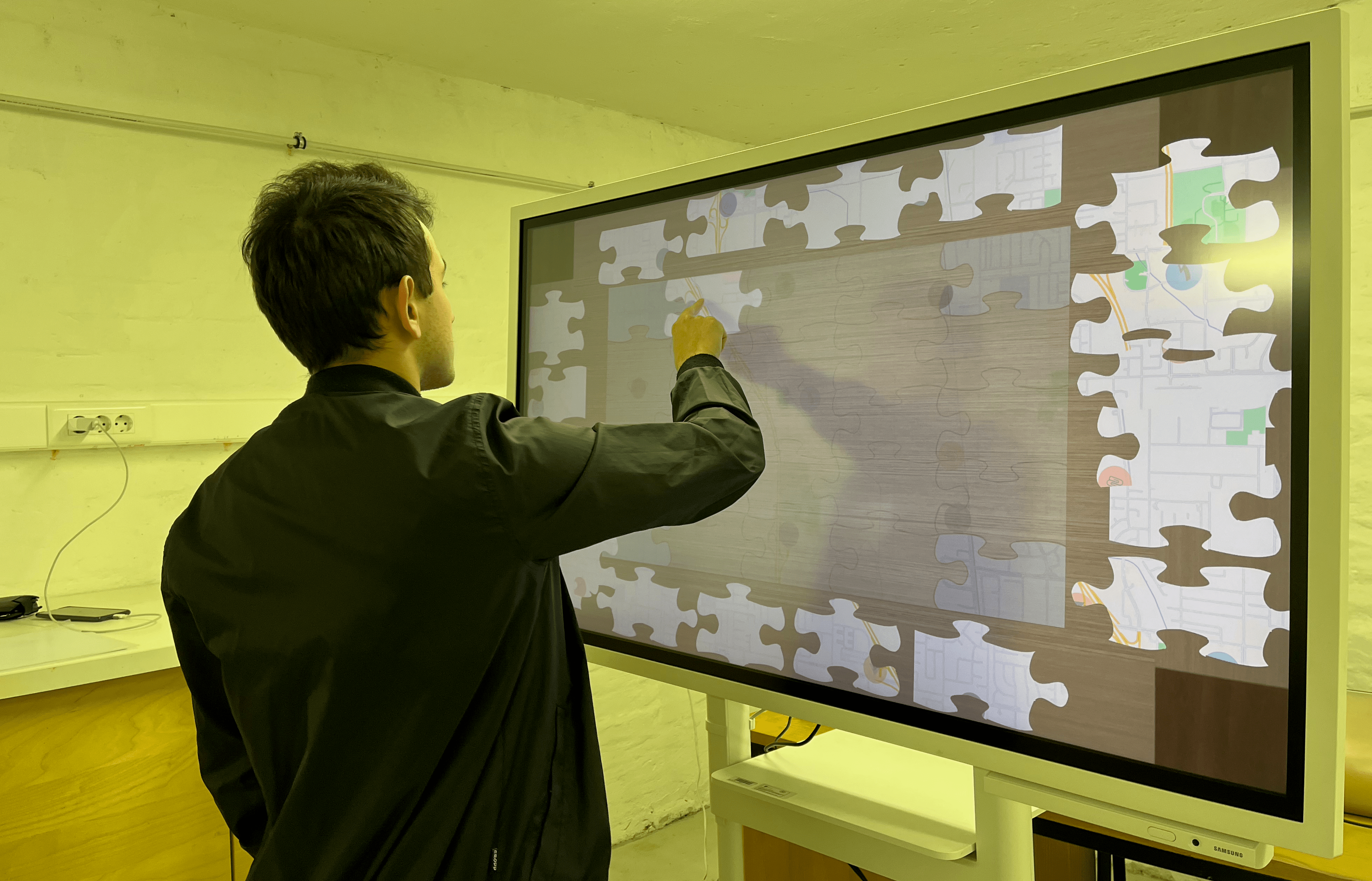}
  \caption{User completing a jigsaw puzzle of a map on the interactive display. When the jigsaw is completed the POIs on the map become interactive.}
   \label{fig:interactivedisplay}
\end{figure}

\par This section discusses the design and concept behind our prototype entitled \textit{Retzzles}. We designed our interaction following the theory of constructivism, according to which individuals actively construct knowledge and create their experiences~\cite{fosnot2013constructivism}. While the end goal of the puzzle is the same for every user, the path toward achieving a solution will be unique for each user. As the puzzle is being built, the user is piecing together the knowledge as long as they play the game. With \textit{Retzzles}, users complete a jigsaw of a tourist destination map individually; however, the final map is the same for everyone. 


\par \textit{Retzzles} is shown on an interactive touch display where users can move the puzzle pieces and put them in the correct locations as seen in \autoref{fig:interactivedisplay}. We used Unity to build this prototype. Jigsaw puzzles were created using GIMP and its pattern feature ``jigsaw''. The map consisted of 20 jigsaw pieces. We opted for this number as it was shown to be not difficult and also not easy to complete the puzzle in the pilot study. The jigsaw puzzles are all organised into one empty parent game object. Each piece is assigned with the \texttt{DragObject} script, allowing it to be moved and dragged anywhere on the screen. This script defines how the puzzle pieces snap into the correct position. Because our pieces are treated as sprites in Unity, we used \texttt{PolygonCollider2D} to handle the physical collisions. The collider’s shape is defined by a \texttt{freeform} edge made of line segments, which is easily adjustable to cover any shape, in our case, any jigsaw puzzle. 

\par
\begin{figure}[hbt]
  \begin{subfigure}[b]{0.49\textwidth}
    \includegraphics[width=\textwidth]{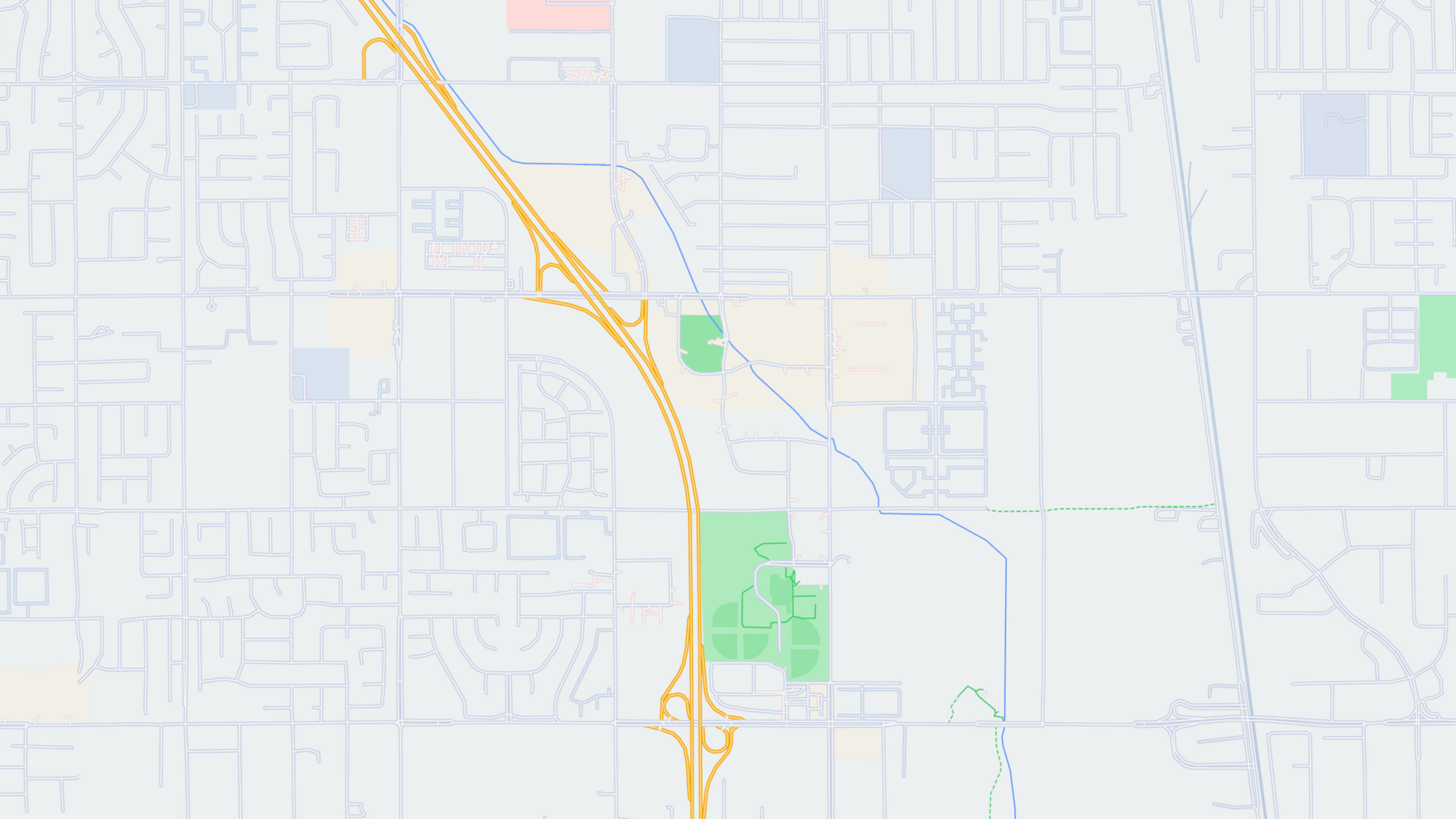}
    \label{fig:training_augmented}
  \end{subfigure}
  \hfill
  \begin{subfigure}[b]{0.5\textwidth}
    \centering
    \includegraphics[width=\textwidth, height=4cm]{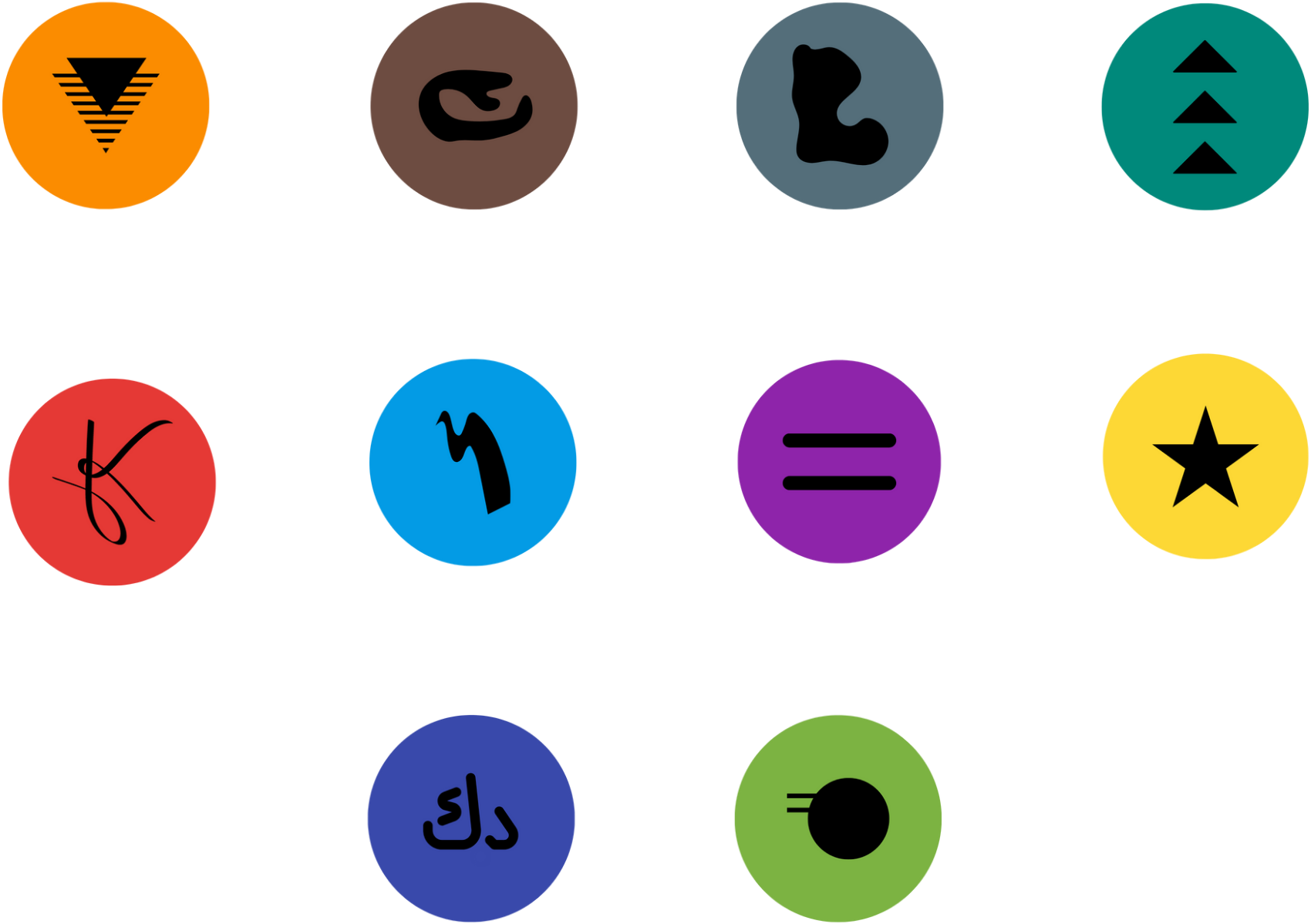}
    \label{fig:lancaster_augmented}
  \end{subfigure}
  \caption{The map and interact-able Points of Interest (POI) of the \textit{Retzzles}: Oxford Suites Hotel, Alderbrook Nursing Facility, Red Salmon Restaurant, U-Haul Moving Supplies, Sun Flower Caffe, Target Grocery Store, Bowlero bowling, Sunnydale School, The Tire Store, Costco Grocery Store and Sunnydale School.}
  \label{fig:mapandpois}
\end{figure}

 When each jigsaw piece is dragged over the correct location and dropped, it snaps automatically in its spot. As the user solves the puzzle, POIs become revealed when the suitable puzzles are pieced together. After completing the jigsaw, the POIs buttons become interactive. Tapping on them reveals additional information about the touristic location they are placed at on the map (\autoref{fig:prototype} bottom right). For this study, we put ten abstract symbols on the map as POIs, as done in~\cite{fujimoto2012relation}. The two reasons behind this decision were to avoid the familiarity with symbols usually used on maps and to not correlate the symbols with the information they revealed when tapped on. The symbols were created using the online graphic design platform Canva\footnote{\url{https://www.canva.com/}}. The map was created using the Snazzy Maps\footnote{\url{https://snazzymaps.com/}} and the ``No label, Bright Colors'' layout as it provided a map with fewer distractions. For the map, we used the city of Lancaster, California, as users were likely not familiar with its layout, as seen in \autoref{fig:mapandpois} left. For the same reason, we used 10 POIs users were likely unfamiliar with.

\par We have added some automated mechanisms to log data within the prototype for more efficient data collection. We tracked the number of moves/clicks for each puzzle piece, activated POIs, calculated the time needed to solve the puzzle, and opened and read the POIs' information. We also recorded and saved every movement of each piece. 

\section{User Study}

For this study, we proposed the following hypothesis: $H_{0}$: Puzzle interactions aid in improved visual, spatial and textual recall of map information. 

\begin{figure}
    \centering
    \begin{subfigure}[b]{0.49\textwidth}
        \centering
        \includegraphics[width=\textwidth]{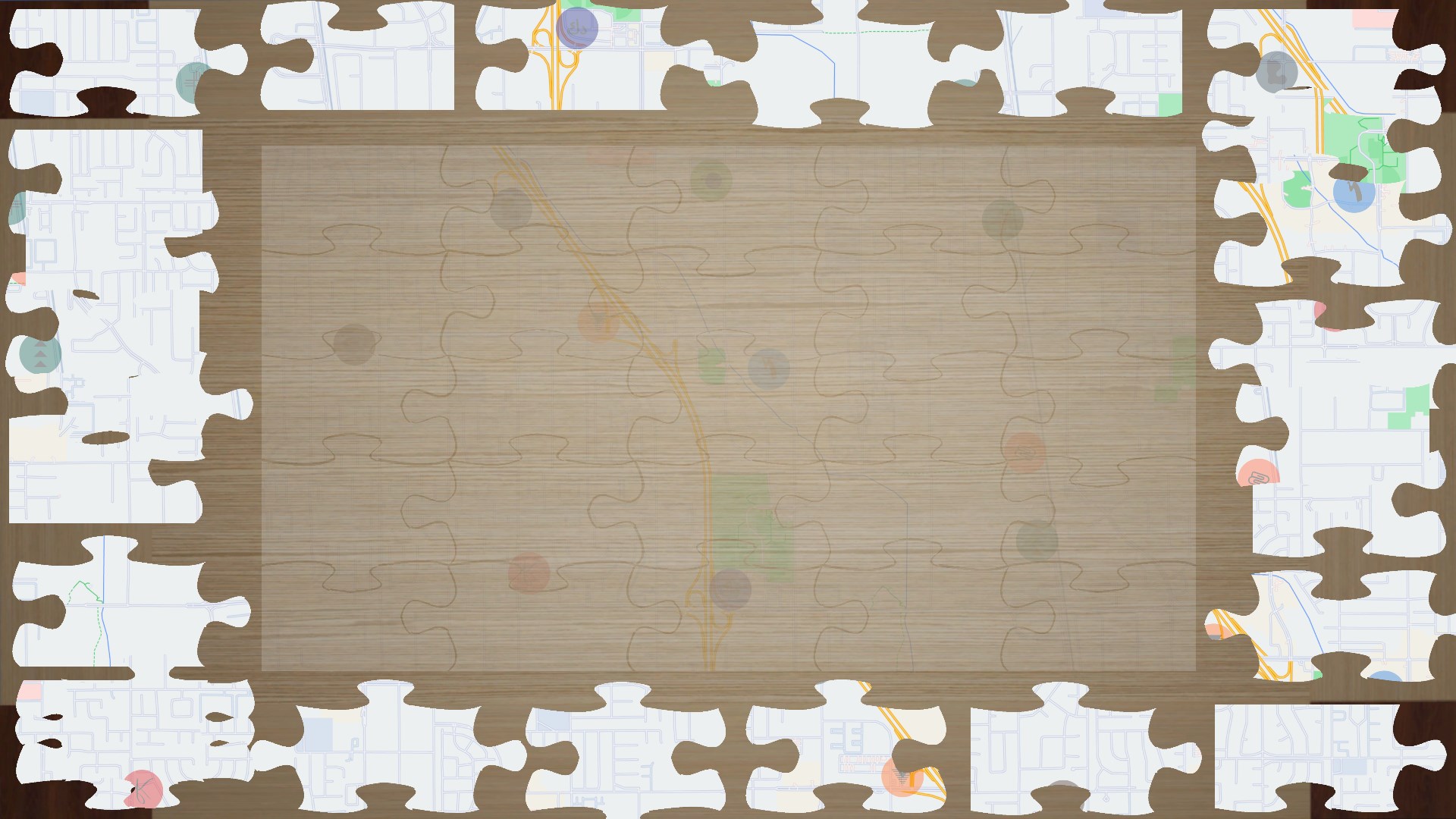}
    \end{subfigure}
    \begin{subfigure}[b]{0.49\textwidth}
        \centering
        \includegraphics[width=\textwidth]{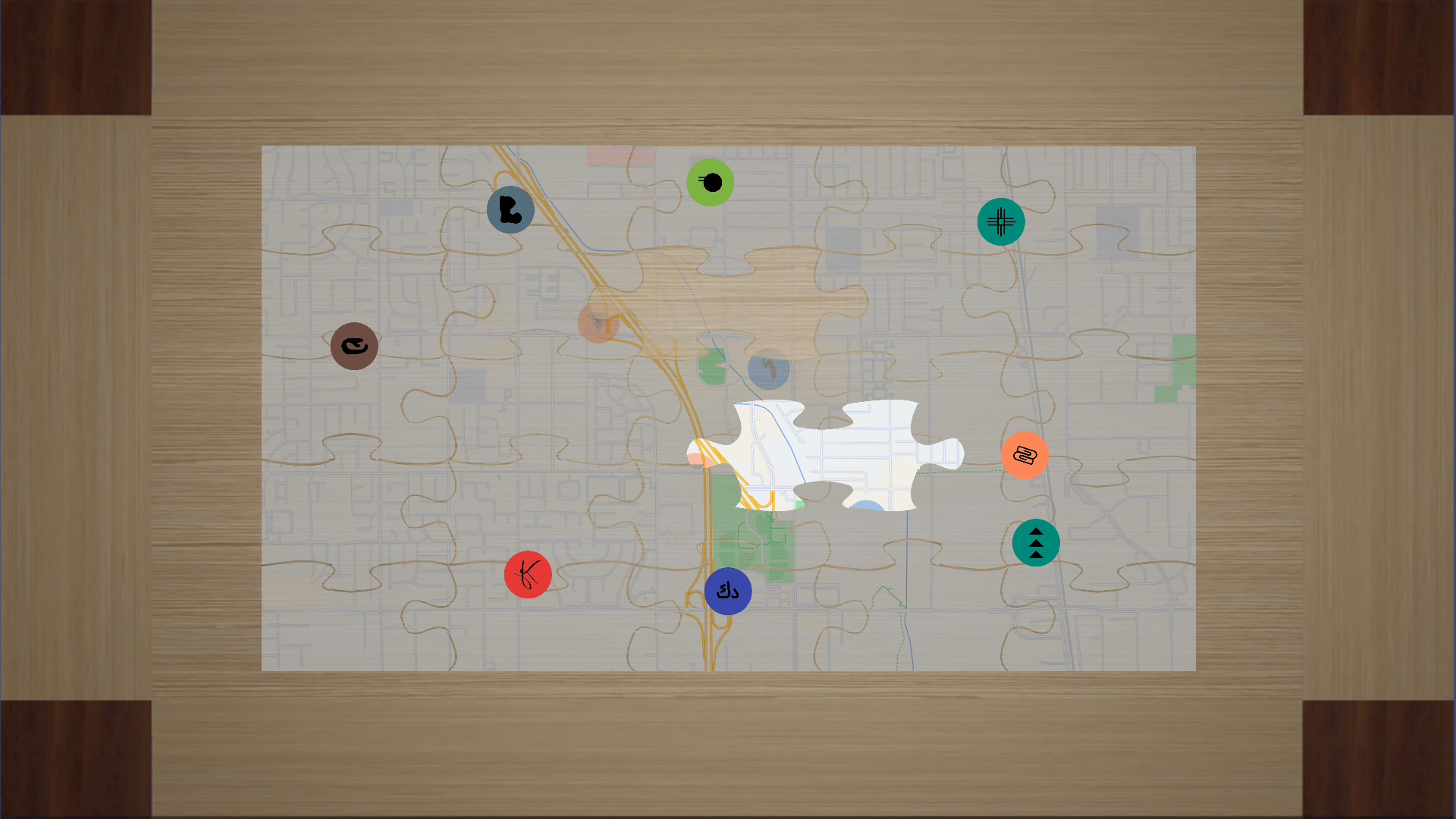}
    \end{subfigure}    
    \vspace*{0.5cm} 
    \begin{subfigure}[b]{0.49\textwidth}
        \centering
        \includegraphics[width=\textwidth]{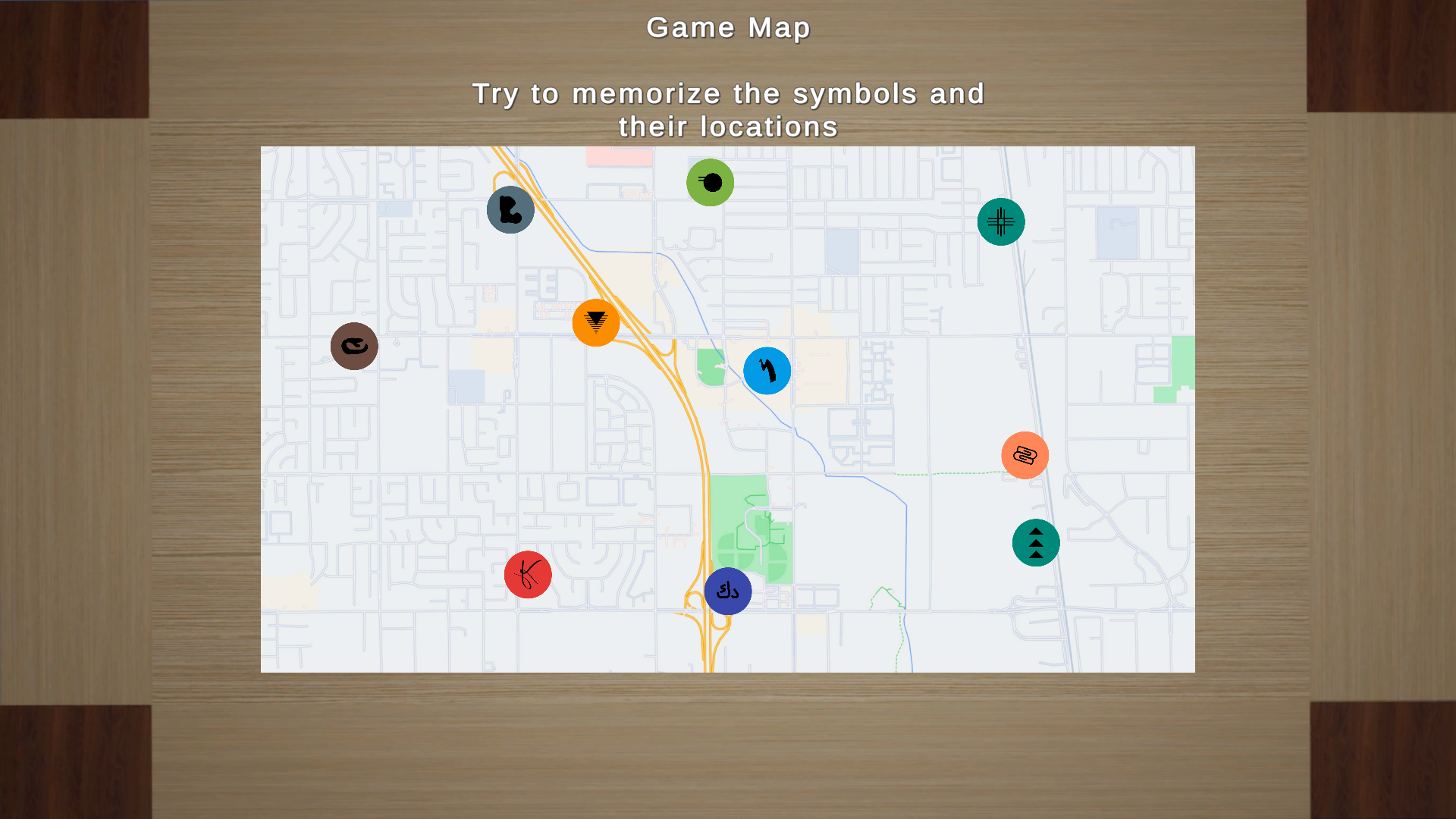}
    \end{subfigure}    
    \begin{subfigure}[b]{0.49\textwidth}
        \centering
        \includegraphics[width=\textwidth]{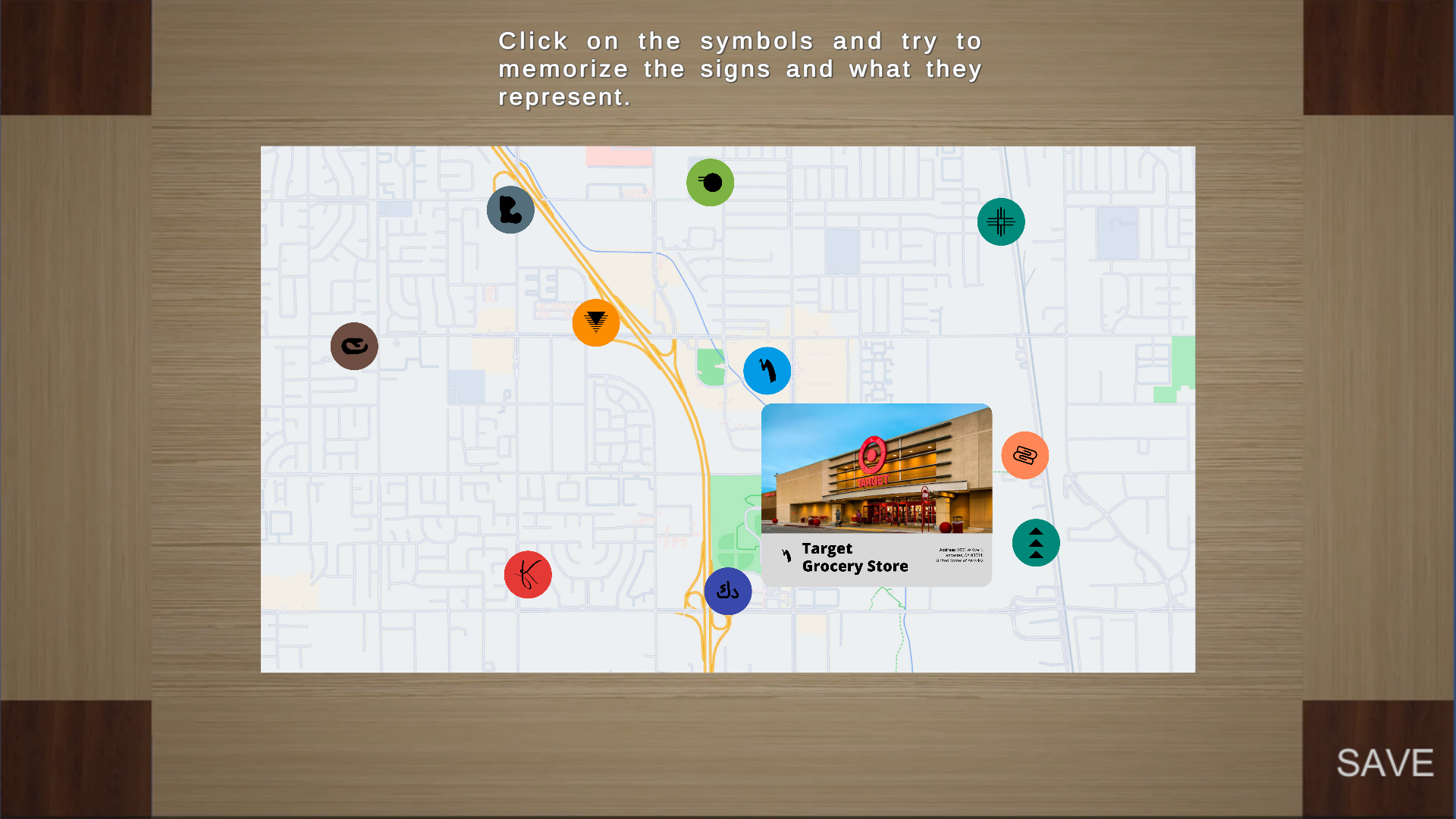}
    \end{subfigure}
    \caption{Different conditions and stages of Retzzles. 
    Top Left: the initial stage of the puzzle condition.  
    Top right: One piece left; when it will be placed in the map will become interactive as in bottom left.
    Bottom Left:  the initial stage of the map-only condition.
    Bottom Right: The information card opened when tapped on the corresponding POI.}
    \label{fig:prototype}
\end{figure}

\par We conducted a between-subject study design to validate our hypothesis with $n=28$ participants recruited through convenience sampling. We recruited participants who either have experience with 2D games or have prior experience in solving puzzles. For the study, an unfamiliar tourist place, such as the city of Lancaster, was used as the subject of the map in the prototype across both conditions. The study involved three phases: 1) Orientation and Informed Consent, 2) Interaction Task and 3) Wrapping phase. In 1), the moderator of the study provided a brief introduction to the experiment and the prototype \textit{Retzzles}. Next, the participants were invited to read and sign the consent form. At the beginning of 2), the moderator explained the task in more detail. We used two conditions, namely (1) Map condition (\autoref{fig:prototype} bottom left)  and (2) Puzzle condition (\autoref{fig:prototype}) top left). Symbols and information used between both conditions were the same. However, the interaction amount differed as the latter condition required puzzle assembly before unlocking the touch-screen map interactions. After interacting with the prototype, participants answered different questionnaires to measure visual, spatial and textual recall.  

\par The aim in both conditions was to memorise as much information as possible after engaging with the prototype. The participants were first given a module that instructed them on utilising the prototype. When they were ready, they proceeded to the actual task based on the conditions set for them. Participants in both conditions were instructed to interact with the prototype while remembering as much information as possible. Interaction for the Map condition consisted of engaging with a map as the POIs were already interactive. Interaction with the Puzzle condition required participants to solve the jigsaw puzzle of a map before being able to interact with the POIs and review the information provided. 

\section{Results and Discussion}

\begin{table}[!t]
\caption{Overview of NASA-TLX Cognitive Scores}
\begin{tabular}{@{}llllrlrlrlrlrlr@{}}
\toprule
\multicolumn{2}{l}{}                                         & \multicolumn{2}{l}{\textbf{cond}} & \multicolumn{2}{r}{\textbf{MD}} & \multicolumn{2}{r}{\textbf{PD}}    & \multicolumn{2}{r}{\textbf{TD}} & \multicolumn{2}{r}{\textbf{P}} & \multicolumn{2}{r}{\textbf{E}} & \textbf{F}           \\ \midrule
\multicolumn{2}{l}{\multirow{2}{*}{\textbf{mean}}}           & \multicolumn{2}{l}{puzzle}        & \multicolumn{2}{r}{\textcolor{green}{\ding{116}}36.1}        & \multicolumn{2}{r}{\textcolor{green}{\ding{116}}8.21}           & \multicolumn{2}{r}{\textcolor{red}{\ding{115}}24.6}        & \multicolumn{2}{r}{\textcolor{green}{\ding{116}}28.6}       & \multicolumn{2}{r}{\textcolor{green}{\ding{116}}37.5}        & \textcolor{green}{\ding{116}}12.1                 \\
\multicolumn{2}{l}{}                                         & \multicolumn{2}{l}{map}           & \multicolumn{2}{r}{49.3}        & \multicolumn{2}{r}{8.57}           & \multicolumn{2}{r}{19.3}        & \multicolumn{2}{r}{29.6}       & \multicolumn{2}{r}{45.0}        & 14.3                 \\
\multicolumn{2}{l}{\multirow{2}{*}{\textbf{std dev}}}        & \multicolumn{2}{l}{puzzle}        & \multicolumn{2}{r}{26.3}        & \multicolumn{2}{r}{05.04}          & \multicolumn{2}{r}{22.5}        & \multicolumn{2}{r}{16.0}       & \multicolumn{2}{r}{25.9}        & 6.71                 \\
\multicolumn{2}{l}{}                                         & \multicolumn{2}{l}{map}           & \multicolumn{2}{r}{19.2}        & \multicolumn{2}{r}{06.02}          & \multicolumn{2}{r}{16.0}        & \multicolumn{2}{r}{18.0}       & \multicolumn{2}{r}{24.3}        & 14.9                 \\
\multicolumn{2}{l}{\multirow{2}{*}{\textbf{Shapiro-Wilk $W$}}} & \multicolumn{2}{l}{puzzle}        & \multicolumn{2}{r}{0.899}       & \multicolumn{2}{r}{0.697}          & \multicolumn{2}{r}{0.792}       & \multicolumn{2}{r}{0.906}      & \multicolumn{2}{r}{0.850}       & 0.879                \\
\multicolumn{2}{l}{}                                         & \multicolumn{2}{l}{map}           & \multicolumn{2}{r}{0.958}       & \multicolumn{2}{r}{0.672}          & \multicolumn{2}{r}{0.785}       & \multicolumn{2}{r}{0.915}      & \multicolumn{2}{r}{0.962}       & 0.697                \\
\multicolumn{2}{l}{\multirow{2}{*}{\textbf{Shapiro-Wilk $p$}}} & \multicolumn{2}{l}{puzzle}        & \multicolumn{2}{r}{0.109}       & \multicolumn{2}{r}{\textless .001} & \multicolumn{2}{r}{0.004}       & \multicolumn{2}{r}{0.139}      & \multicolumn{2}{r}{0.022}       & 0.057                \\
\multicolumn{2}{l}{}                                         & \multicolumn{2}{l}{map}           & \multicolumn{2}{r}{0.686}       & \multicolumn{2}{r}{\textless .001} & \multicolumn{2}{r}{0.003}       & \multicolumn{2}{r}{0.185}      & \multicolumn{2}{r}{0.750}       & \textless .001       \\ \midrule
\multicolumn{2}{l}{\textbf{Mann-Whitney $U$}}                  & \multicolumn{2}{l}{$p$}             & \multicolumn{2}{r}{**0.122}       & \multicolumn{2}{r}{}               & \multicolumn{2}{r}{**0.758}       & \multicolumn{2}{r}{**0.871}      & \multicolumn{2}{r}{**0.309}       & \multicolumn{1}{l}{} \\
\multicolumn{2}{l}{}                                         & \multicolumn{13}{l}{$H_{a}$ $\mu_{puzzle} \neq \mu_{map}$ ,**no sign. diff.}                                                                                                                                                                                                                  \\ \bottomrule
\end{tabular}
\label{tab: results_2}
\vspace{0.5cm}
\footnotesize{\\ Legend: \textit{Cond}: Condition, \textit{MD}: Mental Demand, \textit{PD}: Physical Demand, \textit{D}: Temporal Demand, \textit{P}: Performance, \textit{E}: Effort, \textit{F}: Frustration}\\
\end{table}

\subsection{Information Recall}

\begin{table}[t]
\caption{Overview of Metrics Used. Green arrow meaning better performance compared to the map condition, red arrow meaning worse performance. Legend: \textit{Cond}: Condition, \textit{VR}: Visual Recall, \textit{NS-VR}: Negative Symbol Recall, \textit{SR}: Spatial Recall, \textit{N-TR}: Name Textual Recall, \textit{C-TR}: Category Textual Recall}
\begin{tabular}{@{}llllrlrlrlrlrl@{}}
\toprule
\multicolumn{2}{l}{}                                              & \multicolumn{2}{l}{\textbf{cond}} & \multicolumn{2}{r}{\textbf{VR}}         & \multicolumn{2}{r}{\textbf{NS-VR}}                & \multicolumn{2}{r}{\textbf{SR}}                  & \multicolumn{2}{r}{\textbf{N-TR}}           & \multicolumn{2}{r}{\textbf{C-TR}}       \\ \midrule
\multicolumn{2}{l}{}                                              & \multicolumn{2}{l}{puzzle}             & \multicolumn{2}{r}{{\textcolor{red}{\ding{116}}75.7}}  & \multicolumn{2}{r}{{\textcolor{red}{\ding{116}}7.14}}           & \multicolumn{2}{r}{{\textcolor{green}{\ding{115}}68.6}}  & \multicolumn{2}{r}{{\textcolor{green}{\ding{115}}50.7}}  & \multicolumn{2}{r}{{\textcolor{green}{\ding{115}}68.6}}  \\
\multicolumn{2}{l}{\multirow{-2}{*}{\textbf{mean}}}               & \multicolumn{2}{l}{map}                & \multicolumn{2}{r}{{80.7}}  & \multicolumn{2}{r}{{6.43}}           & \multicolumn{2}{r}{{60.7}}  & \multicolumn{2}{r}{{47.1}}  & \multicolumn{2}{r}{{57.9}}  \\
\multicolumn{2}{l}{}                                              & \multicolumn{2}{l}{puzzle}             & \multicolumn{2}{r}{{18.7}}  & \multicolumn{2}{r}{{8.25}}           & \multicolumn{2}{r}{{26.3}}  & \multicolumn{2}{r}{{25.3}}  & \multicolumn{2}{r}{{11.7}}  \\
\multicolumn{2}{l}{\multirow{-2}{*}{\textbf{std dev}}} & \multicolumn{2}{l}{map}                & \multicolumn{2}{r}{{18.6}}  & \multicolumn{2}{r}{{11.5}}           & \multicolumn{2}{r}{{29.2}}  & \multicolumn{2}{r}{{17.7}}  & \multicolumn{2}{r}{{21.9}}  \\
\multicolumn{2}{l}{}                                              & \multicolumn{2}{l}{puzzle}             & \multicolumn{2}{r}{{0.915}} & \multicolumn{2}{r}{{0.767}}          & \multicolumn{2}{r}{{ 0.880}} & \multicolumn{2}{r}{{0.956}} & \multicolumn{2}{r}{{0.936}} \\
\multicolumn{2}{l}{\multirow{-2}{*}{\textbf{Shapiro-Wilk $W$}}}     & \multicolumn{2}{l}{map}                & \multicolumn{2}{r}{{0.876}} & \multicolumn{2}{r}{{0.638}}          & \multicolumn{2}{r}{{0.882}} & \multicolumn{2}{r}{{0.948}} & \multicolumn{2}{r}{{0.928}} \\
\multicolumn{2}{l}{}                                              & \multicolumn{2}{l}{puzzle}             & \multicolumn{2}{r}{{0.184}} & \multicolumn{2}{r}{{0.002}}          & \multicolumn{2}{r}{{0.059}} & \multicolumn{2}{r}{{0.661}} & \multicolumn{2}{r}{{ 0.370}} \\
\multicolumn{2}{l}{\multirow{-2}{*}{\textbf{Shapiro-Wilk $p$}}}     & \multicolumn{2}{l}{map}                & \multicolumn{2}{r}{{0.051}} & \multicolumn{2}{r}{{\textless .001}} & \multicolumn{2}{r}{{0.062}} & \multicolumn{2}{r}{{0.525}} & \multicolumn{2}{r}{{0.289}} \\ \midrule
\multicolumn{2}{l}{\textbf{Mann-Whitney $U$}}                       & \multicolumn{2}{l}{$p$}                  & \multicolumn{2}{r}{**0.452}                        & \multicolumn{2}{r}{}                                      & \multicolumn{2}{r}{**0.328}                        & \multicolumn{2}{r}{**0.693}                        & \multicolumn{2}{r}{**0.116}                       \\
\multicolumn{2}{l}{}                                             & \multicolumn{12}{l}{$H_{a}$ $\mu_{puzzle} \neq \mu_{map}$ ,**no sign. diff.}                                                                                                                               \\ \bottomrule
\end{tabular}
\label{tab: results}
\end{table}

\par We evaluated visual, spatial, and textual recall and measured participants' subjective cognitive load using the NASA-TLX questionnaire. To test visual recall, participants were presented with 20 abstract symbols (10 symbols were present on the map while the others were not) and asked to identify which 10 were on the map. The results are in the percentage of correct symbols recalled (Visual Recall [VR]) or the percentage of incorrect symbols recalled (Negative Symbol Recall [NS-VR]). Spatial recall was assessed by asking users to remember the positions of the abstract symbols on the map and to place them in their correct locations on a separate blank map [SR]. Finally, for textual recall, users were shown ten symbols on the map and asked to provide information about each POI, including its name (Name Textual Recall [N-TR]) and category (Category Textual Recall [N-TR). For example, they might provide the name ``Costco'' and the category ``grocery store''. 

\par The main goal of the analysis was to compare map and puzzle conditions. We used the Shapiro-Wilk test to explore the normality of data distribution. We found that the data is normally distributed ($p\geq0.05$) for all but the visual recall. We decided to run all statistical tests using the Mann-Whitney U test, a form of independent samples t-test used when the normal distribution requirement is unmet.

\par Based on the participants' mean scores alone (in percentages), the puzzle condition performed better in all metrics except for the visual recall ($SR: 68.6, N-TR: 50.7, C-TR: 68.6$)~\autoref{tab: results}. Likewise, for the cognitive load, we can see that the puzzle condition was less mentally demanded than the map-only condition ($MD: 36.1$, see \autoref{tab: results_2}). We conducted the Mann-Whitney U test, a form of independent samples t-test, to further test the significance of the different metrics.
However, based on our sample size of 28 participants, we found no significant difference ($p\geq0.05$) across all variables. Thus, we rejected our null hypothesis despite better results of the puzzle condition for the spatial and textual recall. 

\par Our sample size may not have been large enough to reach a more conclusive result based on the participants' mean scores in the puzzle condition, which were higher for spatial and textual memory than the map condition. Other factors may have contributed to the effects on visual recall as well. One is that users in the Puzzle condition were using the prototype longer than users in the map condition, which was expected. For example, several participants mentioned associating symbols with the colours used in the POIs, which our study did not consider a variable. 

\subsection{Time and Interaction}

\par As participants used the system, an internal logging mechanism took note of the usage times with the prototype (from opening the scene until solving the puzzle). From here, we collected information such as how much time they interacted with the prototype overall and per puzzle piece. The overall duration from the start of the interaction to completing the condition varied greatly. The shortest duration recorded in the puzzle condition was 157 seconds, while the longest was 778 seconds, with an average of 329.21 seconds. In the map condition, the shortest time was 110 seconds, the longest was 622 seconds, and the average time was 316.35 seconds.

\par Furthermore, the quantity of POI clicks varied, providing a better sense of participant engagement. In the puzzle condition, the lowest number of POI clicks was 21, the maximum number was 82, and the average was 41.31. In the map condition, the lowest number of clicks was 20, the maximum number was 72, and the average was 35.5. 

\subsection{Puzzle Specific Findings}

\par
\begin{figure}
    \centering
    
    \begin{subfigure}[t]{1\textwidth}
        \centering
        \includegraphics[width=0.7\textwidth]{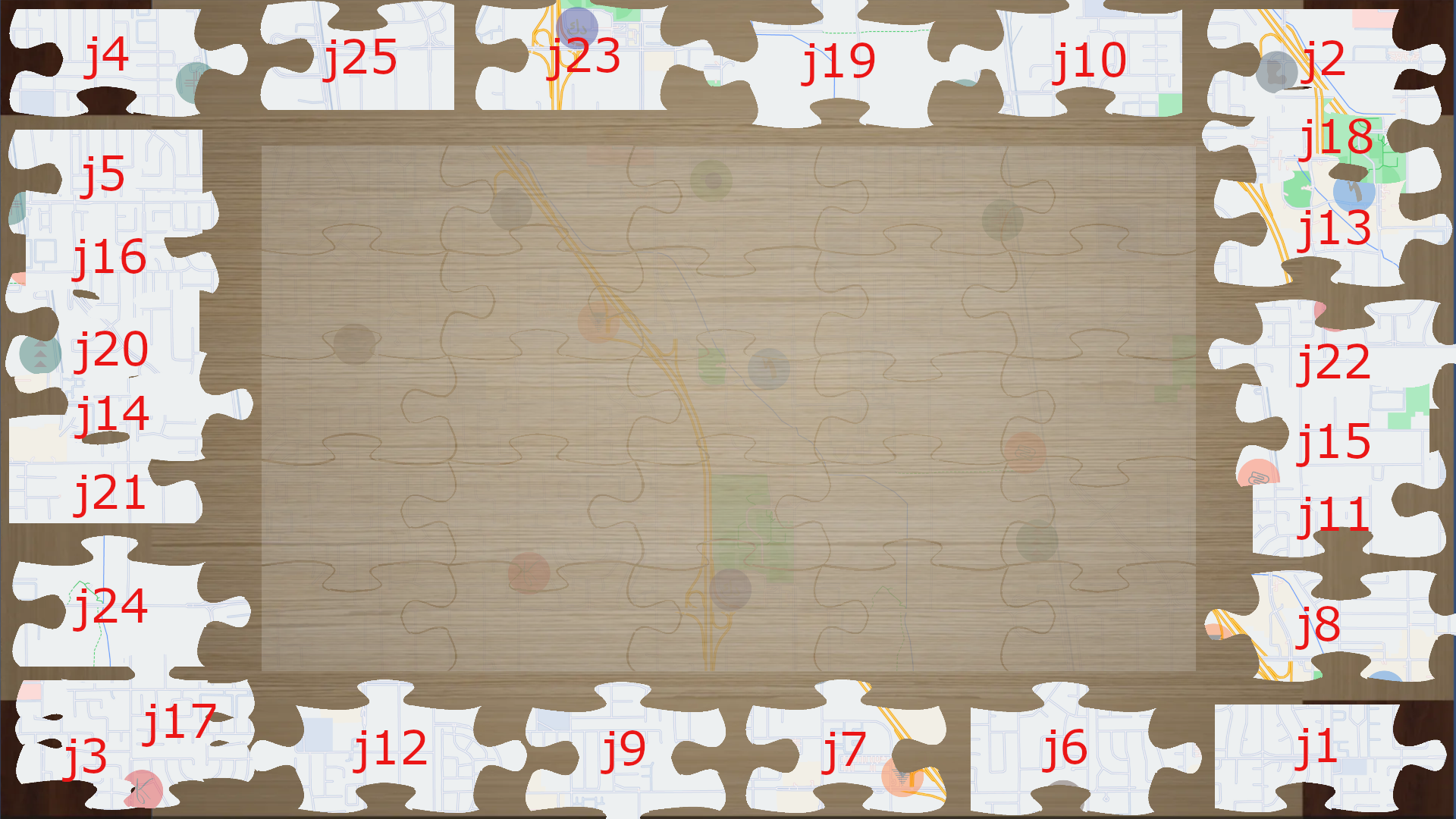}
        \caption{The starting positions of the puzzle pieces.}
        \label{fig:starting_positions}
    \end{subfigure}

    \vspace{0.5cm} 

    \begin{subfigure}[b]{1\textwidth}
        \centering
        \includegraphics[width=0.7\textwidth]{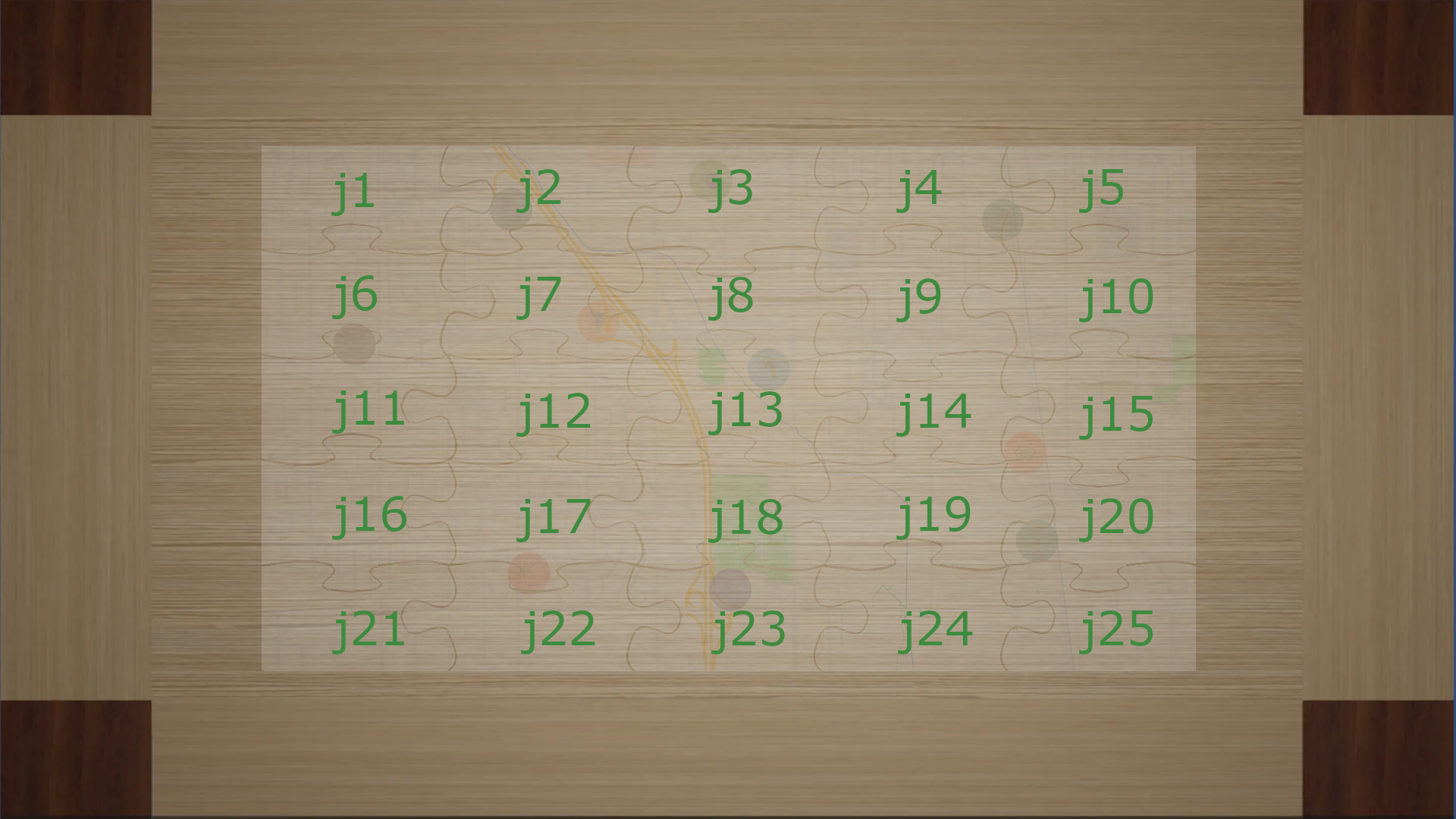}
        \caption{The ending positions of the puzzle pieces.}
        \label{fig:ending_positions}
    \end{subfigure}
    \caption{Look at puzzle positions in the prototype \textit{Retzzles}.}
    \label{fig:puzzle_positions}
\end{figure}

 This section reports on participants' puzzle-solving behaviour. To better understand the following findings a clear view of starting and ending positions of puzzle is needed, which is visible in \autoref{fig:puzzle_positions}.

\begin{itemize}
    \item Patterns in piece completion: We noticed that certain parts were consistently completed by the majority of participants early on, indicating their attractiveness or perceived importance in the puzzle-solving process. These were the pieces: \textbf{j23} (4 out of 14 participants), \textbf{j25} (4 out of 14 participants), and \textbf{j1} (3 out of 14 participants). The parts that took the least time to complete were \textbf{j21} (24 seconds), \textbf{j25} (24 seconds), \textbf{j10} (26 seconds), and \textbf{j1} (30 seconds). The pieces with the fewest pickups were \textbf{j1}, \textbf{j2}, \textbf{j5}, \textbf{j10}, \textbf{j15}, \textbf{j23} and \textbf{j25}, which all have 14 pickups. This indicates that each participant chose them only once when solving the puzzle.

    \item Varied completion times: The fastest completion time was \textbf{98 seconds}. The slowest completion time was \textbf{255 seconds}. This tells us that participants exhibited different levels of efficiency when solving puzzles. On the other hand, the average completion time was: \textbf{133 seconds}.

    \item Troublesome pieces: Certain pieces appeared to cause difficulty for multiple participants, as evidenced by longer completion times or frequent backtracking. These pieces may warrant further investigation for potential design improvements. The most often completed parts were \textbf{j12} (4 out of 14 participants) and \textbf{j9} (3 out of 14 participants). These are the centre parts and have no POI components on them. The sections that took the longest to complete were: \textbf{j14} (99 seconds), \textbf{j4} (81 seconds), \textbf{j19} (77 seconds), and \textbf{j20} (74 seconds). Here we have border pieces \textbf{j4} and \textbf{j20}, as well as middle parts \textbf{j14} and \textbf{j19}. All of them have POI elements on them. We can gain more information into this by examining how many chunk instances of parts exist across all participants. The top three sections in terms of the number of pickups are: \textbf{j14} (24 chunks), \textbf{j17} (23 pickups), and \textbf{j18} (22 pickups).

    \item Participant-specific strategies: Each participant demonstrated distinct puzzle-solving approaches, emphasising individuality and creativity in problem-solving behaviours. Each participant took a different approach to solving the puzzle.
\end{itemize}

\section{Conclusion and Future Work}

\par In this paper, we explored whether engaging with maps in the form of jigsaw puzzle pieces can help users remember more information from the map. We present our prototype, \textit{Retzzles}, used to test this. In one condition, users first had to solve jigsaw puzzles on a map before being able to interact with points of interest displaying essential information. In another condition, the interactive map was already present. Results of our between-subject study, involving $n = 28$ participants, showed better mean scores for spatial and textual recall; however, no significant difference between these conditions was found. Our initial analysis also found no observable differences in the time spent interacting with the puzzle. Despite this, the puzzle condition has shown a slight advantage. While this might be simply because users spent more time with the puzzle, which they found playful, it adds to the knowledge that active learning has advantages. 


\bibliography{main}

\end{document}